\documentclass[useAMS,usenatbib]{mn2e}
\usepackage{graphicx}
\usepackage{rotating}   
\title[Asteroseismology of \mbox{$\theta$ Oph}: spectroscopy]{An asteroseismic study of the $\beta$ Cephei star \mbox{$\theta$ Ophiuchi}: spectroscopic results\thanks{Based on observations obtained with the CORALIE, FEROS and GIRAFFE \'echelle spectrographs attached respectively to the 1.2m Leonard Euler telescope (La Silla, Chile), to the ESO 2.2m telescope (La Silla, Chile) and to the SAAO 1.9m telescope (South Africa).}} 
\author[M. Briquet et al.]{M. Briquet$^{1}$\thanks{E-mail:
maryline@ster.kuleuven.ac.be}\thanks{Postdoctoral Fellow of the Fund for Scientific Research, Flanders}, K. Lefever$^{1}$, K. Uytterhoeven$^{1}$, C. Aerts$^{1}$\\
$^{1}$Instituut voor Sterrenkunde, Katholieke Universiteit Leuven, Celestijnenlaan 200 B, B-3001 Leuven, Belgium\\}

\begin{document}

\date{Accepted 1988 December 15. Received 1988 December 14; in original form 1988 October 11}

\pagerange{\pageref{firstpage}--\pageref{lastpage}} \pubyear{2002}

\maketitle

\label{firstpage} 

\begin{abstract}
We present the results of a detailed analysis of 121 ground-based high-resolution high S/N spectroscopic measurements spread over 3 years for the $\beta$ Cephei star \mbox{$\theta$ Ophiuchi}. We discovered \mbox{$\theta$ Oph} to be a triple system. In addition to the already known Speckle B5 companion of the B2 primary, we showed the presence of a low-mass spectroscopic companion and we derived an orbital period of 56.71 days with an eccentricity of 0.1670. After removing the orbit we determined two frequencies for the primary in the residual radial velocities: $f_1 = 7.1160\ \rm{c\,d}^{-1}$ and $f_2 = 7.4676\ \rm{c\,d}^{-1}$. We also found the presence of $f_3 = 7.3696\ \rm{c\,d}^{-1}$ by means of a two dimensional frequency search across the Si\,III 4567 \AA\ profiles. We identified the $m$-value of the main mode with frequency $f_1$ by taking into account the photometric identifications of the degrees $\ell$. By means of the moment method and the amplitude and phase variations across the line profile, we derived $(\ell_1,m_1) = (2,-1)$. This result allows us to fix the mode identifications of the whole quintuplet for which three components were detected in photometry. This is of particular use for our forthcoming seismic modelling of the primary. We also determined stellar parameters of the primary by non-local thermodynamic equilibrium hydrogen, helium and silicon line profile fitting and we obtained $T_{\rm{eff}} = 24000$ K and $\log g = 4.1$, which is consistent with photometrically determined values.
\end{abstract}


\section{Introduction}
Recently $\beta$ Cephei pulsators have proven to be very promising targets for asteroseismic purposes. Besides the derivation of strong constraints on the mass and metallicity, among other global parameters, the fitting of three independent observed frequencies constrained the core overshooting parameter of the $\beta$ Cephei stars V836 Cen and \mbox{$\nu$ Eri}. Moreover, from the observation of two multiplets, non-rigid rotation was proven for both stars (Aerts et al.\,\cite{aerts2}, Dupret et al.\,\cite{dupret} and Pamyatnykh et al.\,\cite{pamyatnykh}). In addition, it has been shown that non-standard stellar models have to be invoked in order to fit four independent frequencies of \mbox{$\nu$ Eri} (Ausseloos et al.\,\cite{ausseloos}).   

By increasing the number of such asteroseismic studies of additional $\beta$ Cephei target stars, it is clear that our understanding of the internal structure of these massive stars will significantly improve in the near future. In this framework a three-site photometric campaign as well as a long-term spectroscopic monitoring were dedicated to the B\,2\,IV $\beta$ Cephei star \mbox{$\theta$ Ophiuchi} (HD\,157056, V$_{mag}$ = 3.248). The detailed analyses of the gathered photometric and spectroscopic observations are respectively described by Handler et al.\,\citet{handler}, hereafter Paper\,I, and in the present paper. 

In Paper\,I, the frequency analysis and mode identification from 1303 colour photometric measurements show the presence of one radial mode, one rotationally split $\ell=1$ triplet and possibly three components of a rotationally split $\ell=2$ quintuplet. Such a pulsation spectrum is similar to the one observed for V836 Cen (Aerts et al.\,\citet{aerts3}). Consequently fruitful results can be expected from our asteroseismic study devoted to \mbox{$\theta$ Oph}. The present paper describes the detailed analysis of our spectroscopic data with the main aim of identifying the $m$-value of the mode with the highest amplitude, belonging to the quintuplet. In that way we can add an independent well identified mode to the radial one and the central mode of the triplet already detected. Such an identification will allow seismic modelling of the star, which will be done in the near future.

The spectroscopic variability of \mbox{$\theta$ Oph}, as well as the main period of some 0.14 days, have been known for a long time but no spectroscopic mode identification was yet attempted. Radial velocity variations with a main period of 0.2862\,d were first discovered by Henroteau\,(\citet{henroteau}). From new radial velocity measurements, Van Hoof et al.\,\citet{vanhoof1} suggested a period of 0.15\,d, which corresponds to about half Henroteau's period. They also noticed that the mean velocity at the time of their observations was $-8\ \rm{km\ s}^{-1}$, a value distinctly different from Henroteau's values of 1920 and 1922, which were 0 and $-15\ \rm{km\ s}^{-1}$, respectively. McNamara\,\citet{mcnamara} and afterwards Van Hoof and Blaauw \citet{vanhoof2} confirmed the period found by Van Hoof et al.\,\citet{vanhoof1} and refined it to 0.1404\,d. 

From their radial velocities measurements of members of the Sco-Cen association Levato et al.\,\citet{levato} suspected \mbox{$\theta$ Oph} to be a single-lined spectroscopic binary but they could not derive orbital elements. Moreover, McAlister et al.\,\citet{mcalister} detected \mbox{$\theta$ Oph} to be a Speckle binary and Handler et al.\,\citet{handler} determined that the Speckle companion must be a B5 main sequence star. 

In the literature, values for the  projected rotational velocity $v \sin i$ of \mbox{$\theta$ Oph} were derived. Brown and Verschueren\,\citet{brown} measured $v \sin i = 31 \pm 3\ \rm{km\ s}^{-1}$ and Abt et al.\,\citet{abt} found $v \sin i \simeq 30\ \rm{km\ s}^{-1}$ for our studied star, compatible with the previous result. 
 
The paper is organized as follows. In Section\,2 we give a description of our observations and data reductions. In Section\,3 we show that the primary has a low-mass spectroscopic companion besides the distant Speckle companion and we derive orbital parameters. Section\,4 is devoted to the frequency analysis on our data after removing the orbit, followed by the identification of the wavenumbers $(\ell,m)$ for the main frequency $f_1 = 7.1160\ \rm{c\,d}^{-1}$. In Section\,5 we derive the position of \mbox{$\theta$ Oph} in the Hertzsprung-Russel diagram. In Section\,6 we end the paper with a summary.

\section[]{Observations and data reductions}
Our spectroscopic data were obtained with the CORALIE \'echelle spectrograph attached to the 1.2m Leonard Euler telescope in La Silla in Chile, during several runs spread over 2000-2003. We also have at our disposal some measurements gathered with the FEROS and GIRAFFE \'echelle spectrographs attached respectively to the ESO 2.2m telescope (La Silla, Chile) and to the SAAO 1.9m telescope (South Africa). The number of observations and the ranges of their Julian Dates are given in Table\,\ref{log}. The two last runs are respectively a FEROS run and a GIRAFFE run.

\begin{table}
\caption{Observing logbook of our spectroscopy of \mbox{$\theta$ Oph}}
\begin{center}
\begin{tabular}{cccc}
\hline\\[-5pt]
Instrument & Number of & \multicolumn {2}{c}{JD} \\
& observations & \multicolumn {2}{c}{2450000 +}   \\
& & Start & End \\[5pt]
\hline\\[-5pt]
CORALIE & 6 & 1760  & 1761  \\
CORALIE & 5 & 1816  & 1816  \\
CORALIE & 29 & 2015 & 2026 \\
CORALIE & 14 & 2114 & 2123  \\
CORALIE & 32 & 2727 & 2740  \\
FEROS & 8 & 2770 & 2773  \\
GIRAFFE & 27 & 2780 & 2784  \\[5pt]
\hline
\end{tabular}
\end{center}
\label{log}
\end{table}

An on-line reduction of the CORALIE spectra, using the INTER-TACOS software package, is available. For a description of the reduction process we refer to Baranne et al.\,\citet{baranne}. We did a more precise correction for the pixel-to-pixel sensitivity variations by using all available flatfields obtained during the night instead of using only one flatfield, as is done by the on-line reduction procedure. The FEROS data reduction was performed using the on-line FEROS reduction pipeline which makes use of the ESO-MIDAS software package. This reduction includes consecutively the extraction of the inter-order background of the \'echelle spectra, the straightening and extracting of the orders, the removal of the blaze function and the pixel-to-pixel variations, the rebinning of the orders to the wavelength scale by using ThAr calibration spectra and a merging of the \'echelle orders. We performed an additional correction for the wavelength sensitivity of the shape of the internal flatfields by means of a smoothed average of dome flatfields. The GIRAFFE spectra were reduced by using the GIRAFFE pipeline reduction program XSPEC2, which follows a similar procedure as the FEROS pipeline reduction. The spectra of a ThAr arc lamp, taken at regular intervals to calibrate possible drifts in wavelength, and two different types of flatfields were used for the reduction. We note that bias frames were not separately taken as the bias value was calculated from the overscan region on the CCD. Flatfielding was accomplished by using camera flats, which were obtained by illuminating the camera with uniform light using a tungsten filament lamp and a diffusing screen. The blaze correction was determined by measuring the response across each order when the fibre was illuminated by the tungsten lamp (fibre flats). 

Finally, all spectra were normalized to the continuum by a cubic spline function, and the heliocentric corrections were computed.

\begin{figure*}  
\includegraphics[bb=50 50 820 350, width=14cm]{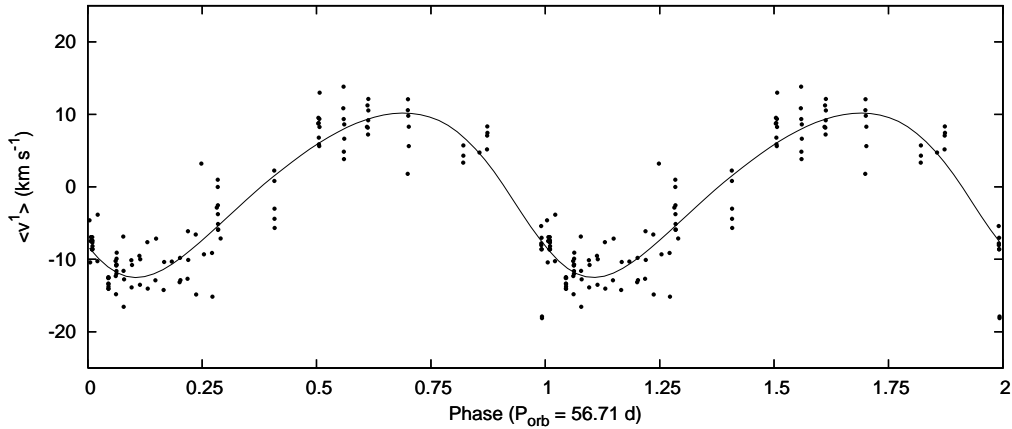}  
\caption{Phase diagram of the radial velocity placed at average zero with respect to the orbital period $P_{\rm{orb}} = 56.71$ d. The line corresponds to the orbital solution obtained with FOTEL (Table\,\ref{elements}).}  
\label{orbit}  
\end{figure*}  

\section[]{A triple system}

For our study of the line-profile variability we used the Si\,III triplet around 4567 \AA. We point out that an effective temperature of some 18400 K was derived in Paper\,I for the B5 Speckle companion, which implies that this companion contributes very little to the Si\,III lines. Moreover, the astrometric data presented by McAlister et al.\,\citet{mcalister} and by
Shatsky \& Tokovinin\,\citet{shatsky} point towards an orbital period of order 100 yr, so our
approximation that it shows no radial-velocity variation on a timescale of 3
years seems justified.

We computed the first three velocity moments \mbox{$<v^1>$}, \mbox{$<v^2>$} and \mbox{$<v^3>$} (see Aerts et al.\,\citet{aerts1} for a definition of the moments of a line profile) of the Si\,III 4553 \AA\ line with the aim of performing a frequency analysis. The integration boundaries of the moments were dynamically determined by visual inspection of each spectrum in order to avoid noisy continuum. To perform the frequency analysis on the first moment \mbox{$<v^1>$}, which is the radial velocity placed at average zero, we used the PDM method (Stellingwerf\,\citet{stellingwerf}) and the Scargle method (Scargle\,\citet{scargle}). We tested frequencies from 0 to 10 cycles per day (\mbox{c d$^{-1}$}) with a frequency step of $0.0001\ \rm{c\ d}^{-1}$. The error estimate of our determined frequencies is between 0.0001 \mbox{c d$^{-1}$} and 0.001 \mbox{c d$^{-1}$}. We obtained the same results with both methods.

The frequency that dominates the radial velocity is 0.0175 \mbox{c d$^{-1}$}. Such a long period of some 57 days does not belong to the pulsation period range of $\beta$ Cephei stars and is too long to be the rotation period of a star with $v \sin i \simeq 30\ \rm{km\ s}^{-1}$ and $R \simeq 5\ R_{\odot}$. Consequently we attributed this variability with a peak-to-peak amplitude of $\simeq 25\ \rm{km\ s}^{-1}$ to binarity. 

In order to derive the orbital parameters of the binary system we used the code FOTEL developed by Hadrava\,\citet{hadrava}. The program FOTEL is designed to solve either simultaneously or separately light- and radial velocity curves of binary stars with a possible distant third component. For \mbox{$\theta$ Oph} we only used radial velocity measurements because there is no sign of the binarity in photometric measurements at our disposal in Paper\,I. We found an orbital period of $P_{\rm{orb}} = 56.71$ days with eccentricity $e = 0.1670$. The other orbital elements and their standard errors are listed in Table\,\ref{elements}. The phase diagram with respect to $P_{\rm{orb}}$ is shown in Fig.\,\ref{orbit}. 

If we consider that the orbital inclination $i_{\rm{orb}}$ corresponds to the inclination of the primary $i_{\rm{rot}}$, the result of the mode identification described in Section\,4.2 leads to $i_{\rm{orb}}$ in the interval [70$^{\circ}$,90$^{\circ}$]. With a mass for the primary component of some 9 $M_\odot$ (see Paper\,I), we conclude that the secondary component has a mass lower than 1 $M_\odot$. We note that the duty cycle of the photometric data in Paper\,I is not sufficient to detect a possible eclipse. 

We therefore conclude that \mbox{$\theta$ Oph} is a triple system composed of a B2 primary, a low-mass spectroscopic secondary and a Speckle B5 star. 

\begin{table}
\caption{The orbital parameters for \mbox{$\theta$ Oph}. $P_{\rm{orb}}$ is the orbital period, $T_o$ the time of periastron passage, $e$ the eccentricity, $K_1$ the semi-amplitude of the radial velocity curve of component 1, $\omega$ the periastron longitude, $v_\gamma$ the system velocity.}
\begin{center}
\begin{tabular}{ccc}
\hline\\[-5pt]
$P_{\rm{orb}}$ (days)     & 56.712   $\pm$ 0.046 \\
$T_o$ (HJD)               & 51816.99 $\pm$ 2.71 \\
$e$                       & 0.1670   $\pm$ 0.043\\
$K_1$ (km s$^{-1}$)       & 11.345   $\pm$ 1.099\\
$\omega$ (degrees)        & 128.046  $\pm$ 0.319\\
$v_\gamma$ (km s$^{-1}$)  & 7.202    $\pm$ 0.369\\
$a_1 \sin i_{\rm{orb}}$ (a.u.)             & 0.058311\\
$f_1(m)$ ($M_\odot$)                & 0.0082243 \\
$rms$ (km s$^{-1}$)       & 3.35 \\[5pt]
\hline
\end{tabular}
\end{center}
\label{elements}
\end{table}

\section[]{A pulsating primary}
While the tertiary contributes significantly to the spectrum of the triple
system, its Si\,III lines are so weak that they must be situated completely
within those of the primary. We have no information about possible variability
of the tertiary. In any case, should it be a rotational or a pulsational
variable star, we expect its frequency range to be vastly different from the one
of the primary, given its mid-B spectra type. Its presence will therefore not
affect the pulsational interpretation of the primary we present below. A similar
situation of having a companion spectral line within the one of the primary
occurs for the $\beta\,$Cephei star $\beta\,$Crucis (Aerts et al.\,\citet{aerts4}).

\subsection{Frequency analysis}
\begin{figure}  
\includegraphics[bb=100 50 600 700, width=9cm]{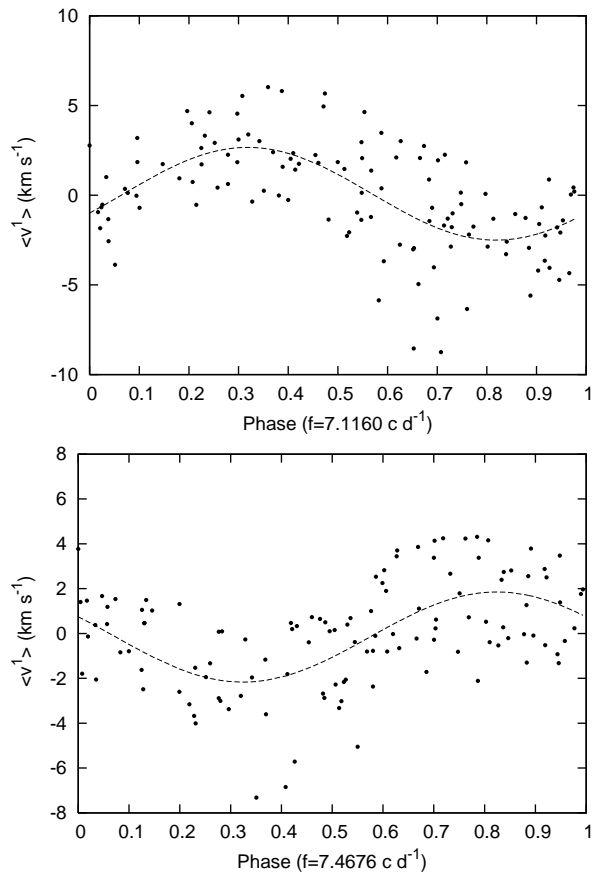}
\caption{Upper panel: phase diagram of the first moment computed from the Si\,III 4553 \AA\ line, after removing the orbit, for $f_1$=7.1160 \mbox{c d$^{-1}$}. Lower panel: phase diagram of the data prewhitened with $f_1$ for $f_2=7.4676$ \mbox{c d$^{-1}$}.}  
\label{v1}  
\end{figure}

After removing the orbit we searched for additional frequencies in the residual radial velocities by subsequent prewhitening. We found $f_1 = 7.1160\ \rm{c\,d}^{-1}$ and $f_2 = 7.4676\ \rm{c\,d}^{-1}$. Phase diagrams for the first moment are shown in Fig.\,\ref{v1}. Note that $f_1$ is the frequency already detected in previous spectroscopic observations of \mbox{$\theta$ Oph} and both frequencies correspond to those found in photometric measurements in Paper\,I. In the radial velocity and in the higher order moments we could not find any additional frequency after prewhitening with $f_1$ and $f_2$.  

\begin{figure*} 
 \includegraphics[bb=370 50 790 600,width=7cm]{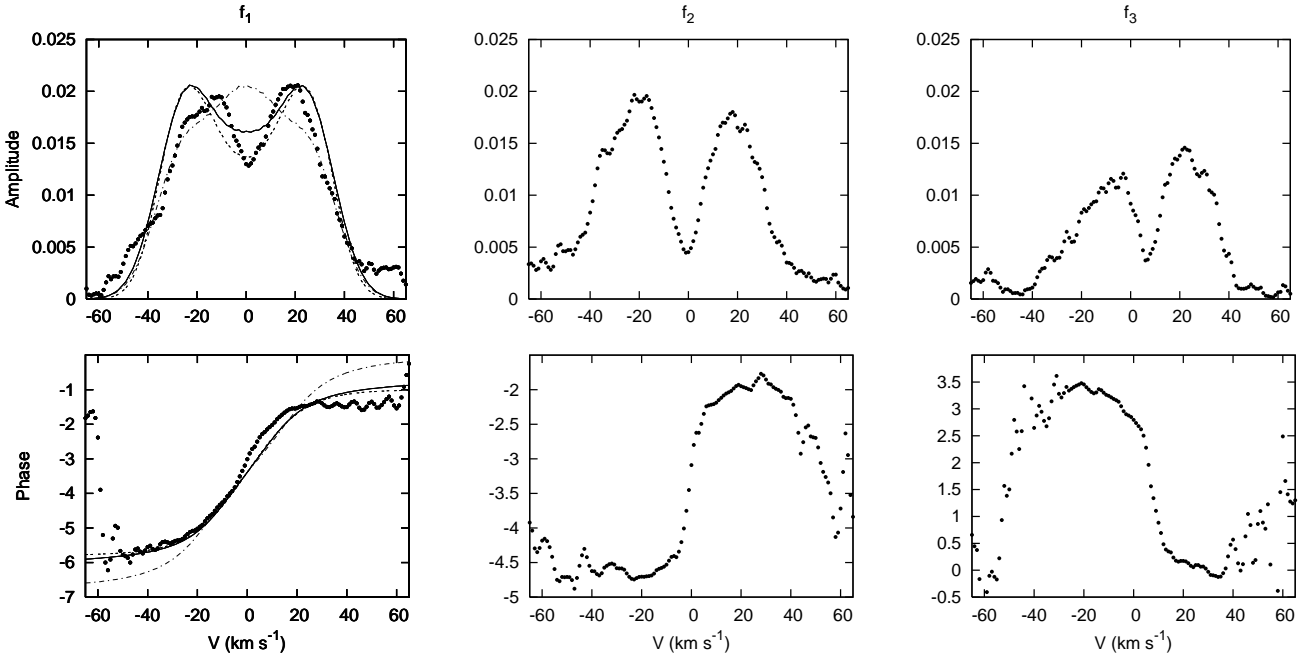}
\caption{Amplitude and phase distributions for $f_1$, $f_2$ and $f_3$ for the Si\,III 4553 \AA\ line (dots) and comparisons for $f_1$ with the theoretical ones for the solutions given in italic in Table\,\ref{mom}. The solutions with $(\ell_1,m_1) = (2,-1)$, $(1,-1)$ and $(2,-2)$ are respectively represented by a full line, a dashed line and a dot-dashed line. The amplitudes are unitless and the phases are expressed in $\pi$ radians.}  
\label{amp_phase}  
\end{figure*}

We then attempted to determine other frequencies by means of a two dimensional Scargle frequency analysis on the spectra themselves. After prewhitening with the main frequency $f_1$ clearly present in the data, we found the presence of both frequencies $f_2$ and $f_3 = 7.3696\ \rm{c\,d}^{-1}$. The latter was also observed in photometry of Paper\,I. The amplitude and phase variations across the Si\,III 4553 \AA\ line are shown in Fig.\,\ref{amp_phase} for $f_1$, $f_2$ and $f_3$.

\subsection{Mode identification}
In Paper\,I three frequencies of a rotationally split $\ell=2$ quintuplet were observed for \mbox{$\theta$ Oph}, which are $7.1160\ \rm{c\,d}^{-1}$, $7.2881\ \rm{c\,d}^{-1}$ and  $7.3697\ \rm{c\,d}^{-1}$. The former and latter correspond to $f_1$ and $f_3$. Consequently the main $\ell = 2$ mode with frequency $7.1160\ \rm{c\,d}^{-1}$ has an $m$-value equal to $-2$ or $-1$. By means of our spectroscopic data we will lift this ambiguity.

To perform our mode identification by different methods we fixed certain parameters while others are fully free. We took a linear limb-darkening coefficient $u$ of 0.32 (see e.g. Wade \& Rucinski \citet{wade}). We used the $K$-value given by $K=GM/\omega^2 R^3$, where $M$ is the mass, $R$ the radius and $\omega$ the angular pulsation frequency. Note that slightly different values of these parameters within the errors do not change the result of the mode identification. In order to define a fine grid of unknown stellar and pulsation parameters without testing useless cases, the velocity amplitude $A_p$ was varied so that the theoretical amplitudes of the first moment correspond to the observed ones within the errors.

\begin{table*}
\caption[]{The ten best solutions of the mode identification through the discriminant $\Sigma$ (see Briquet \& Aerts \citet{briquet}), using the Si III 4553 \AA\ line. We took $K_1 = 0.120$ and $K_2 = 0.109$ for the ratios of the amplitude of the horizontal and of the vertical motion. $A_{p}$ is the amplitude of the radial part of the pulsation velocity, expressed in km\,s$^{-1}$; $v_{\rm{r,max}}$ and $v_{\rm{t,max}}$ are respectively the maximum radial and tangential surface velocity due to the two modes, expressed in km\,s$^{-1}$; $i_{\rm{rot}}$ is the inclination angle, expressed in degrees; $v \sin i$ is the projected rotational velocity, expressed in km\,s$^{-1}$ and $\sigma$ is the intrinsic line-profile width, also expressed in km\,s$^{-1}$.}
\centering
\begin{tabular}{c|ccccccccccc}
\hline\\[-5pt]
($\ell_1$,$m_1$) & $\it{(2,-1)}$  & $\it{(1,-1)}$ & $(3,-1)$ & $(2,0)$ & $(0,0)$ & $(1,0)$  & $\it{(2,-2)}$ & $(3,0)$ & $(1,1)$ & $(2,1)$ &$\ldots$ \\
($\ell_2$,$m_2$) & $\it{(0,0)}$ & $\it{(0,0)}$ & $(0,0)$ & $(0,0)$ & $(0,0)$ & $(0,0)$ & $\it{(0,0)}$ & $(0,0)$ & $(0,0)$ & $(0,0)$ &$\ldots$ \\
$A_p^1$ & \it{33.96} & \it{12.65} & 46.28 & 20.80 & 13.11 & 8.98 & \it{16.98} & 47.43 & 12.65 &38.73  &$\ldots$ \\
$A_p^2$ & \it{10.74}  & \it{10.74}  &  10.74 &  10.74 &  10.74 &  10.74 &  \it{10.74} &  10.74 &  10.74 &  10.74 &$\ldots$ \\
$v_{\rm{r,max}}$ & 3.23 & 7.23 & 17.46 & 9.34 & 6.58 & 3.05 & 9.36 & 3.39 & 7.24 & 3.21 & $\ldots$  \\
$v_{\rm{t,max}}$ & 3.28 & 0.55 & 1.86 & 0.04 & 0.00 & 0.53 & 1.70 & 6.39 & 0.51 & 3.47 & $\ldots$ \\
$i_{\rm{rot}}$ &  \it{75} & \it{89} &  90 & 90 & - & 5 & \it{90} & 69  & 90 & 77 &$\ldots$\\
$v \sin i$ & \it{31} &\it{33.5}  &30  & 38 & 38.5 & 38.5 & \it{31.5} &30  & 32 &30.5 &$\ldots$\\
$\sigma$ & \it{10} & \it{8.5} & 10 &2.5  & 1 & 1 &  \it{10} & 9.5 & 10 &10 &$\ldots$\\
$\Sigma$ & \it{14.40} & \it{14.43} & 14.54 & 14.55 & 14.56 &14.57  & \it{14.77} & 14.77 & 15.01 & 15.13 &$\ldots$\\[3pt]
\hline
\end{tabular}
\label{mom}
\end{table*}

\subsubsection{The moment method}
We first used the moment method. With this method, the wavenumbers $(\ell,m)$ and the other continuous velocity parameters are determined in such a way that the theoretically computed first three moment variations of a line profile best fit the observed ones. We refer to Briquet \& Aerts \citet{briquet} for the latest version of the technique, which was improved by these authors in order to perform a simultaneous identification of all the modes that are present in the data. Since we only detected $f_1 = 7.1160\ \rm{c\,d}^{-1}$ and $f_2 = 7.4676\ \rm{c\,d}^{-1}$ in the first three moments and since $f_2$ is unambiguously identified as a radial mode by photometry (Paper\,I), we performed a multiperiodic mode identification with these two latter frequencies, the mode with frequency $f_2$ being fixed to $(\ell_2,m_2) = (0,0)$. According to the criterion given by Briquet \& Aerts\,\citet{briquet}, we only needed to test $\ell_1 \le 3$.

\begin{figure}  
\includegraphics[bb=60 50 300 500, width=8.6cm]{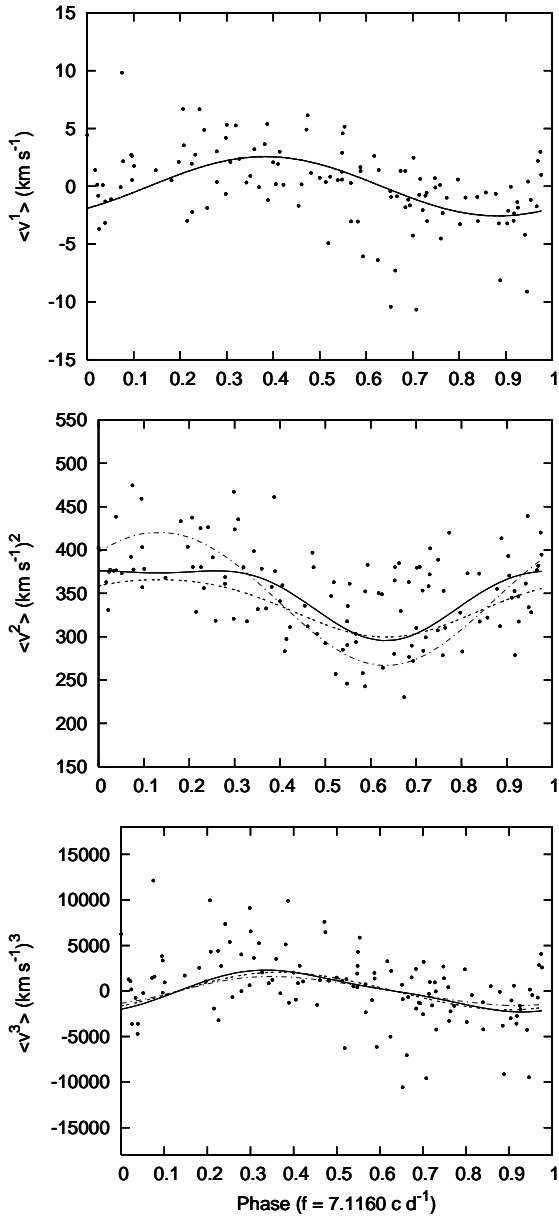}
\caption{Comparisons between the observed first three moments (dots) and three of the best solutions given in italic in Table\,\ref{mom}. The solutions with $(\ell_1,m_1) = (2,-1)$, $(1,-1)$ and $(2,-2)$ are respectively represented by a full line, a dashed line and a dot-dashed line.}  
\label{moment}  
\end{figure}

The results of the mode identification are given in Table\,\ref{mom}. The lower the discriminant $\Sigma$, the better the agreement between theoretical and observed moment values. We can conclude that the mode identification by means of the moment method favours an $m = -1$. Since the photometric mode identification described in Paper\,I excludes an $\ell_1 = 3$, the mode identification gives a preference to $(\ell_1,m_1) = (2,-1)$ or $(1,-1)$. However, the solution $(2,-2)$ remains possible. For the solutions in italic in Table\,\ref{mom}, we compare the moment values with the observed ones in Fig.\,\ref{moment} in order to check if we can choose between these three possibilities. We conclude that we cannot discriminate between $(\ell_1,m_1) = (2,-1)$ or $(1,-1)$ and the solution $(2,-2)$ gives a theoretical second moment amplitude slightly too large compared to the observed one. We also computed observed and theoretical moments up to order 6 in order to attempt to discriminate between the different solutions, as could be done for several slowly pulsating B stars by De Cat et al.\,(\citet{decat}). The solutions with $(2,-1)$ and $(1,-1)$ give very similar first six moments and the solution with $(2,-2)$ shows a slightly less good fit to the even moments.

\subsubsection{The amplitude and phase variations across the line profile}
In order to discriminate between the best moment solutions we also used the amplitude and phase distributions shown in Fig.\,\ref{amp_phase} (see Telting and Schrijvers\,\citet{telting}, Schrijvers and Telting\,\citet{schrijvers} for a definition). Because of the zero amplitude at the center of the line and a corresponding phase shift of $\pi$, one can directly conclude that the mode corresponding to $f_2$ is axisymmetric. It corroborates the photometric mode identification. Moreover, because of opposite slopes for the phase distribution, the modes corresponding to $f_1$ and $f_3$ have $m$-values of opposite sign. We point out that this conclusion can be made for $\beta$ Cephei stars having p-mode pulsations while it cannot be concluded for SPBs with g-modes (De Cat et al.\,\citet{decat}).

In Fig.\,\ref{amp_phase} we also compare the theoretical and observed amplitude and phase variations across the line profile for $f_1$ for the best solutions of the moment method given in italic in Table\,\ref{mom}. This comparison allows us to exclude $(\ell_1,m_1) = (2,-2)$ since the theoretical variations do not mimic the observed ones for this solution.
  
As an additional check we also performed an independent mode identification by means of the following procedure. By using Townsend's \citet{townsend} codes, called BRUCE and KYLIE, we generated theoretical line profile time series for several wavenumbers $(\ell,m)$ and for a large grid of other continuous parameters. Subsequently we computed the amplitude and phase variations across the line profile for each generated line profile time series. Finally, by comparing these theoretically computed amplitude and phase distributions with the observed ones, we derived the couple $(\ell,m)$ and other parameters which led to the best fit. 

Because generating line-profile variations is very time consuming, even with nowadays computers, we performed this procedure for the frequency $f_1$ and for couples $(2,-1)$, $(1,-1)$  and $(2,-2)$ since our goal is to discriminate between these possible $(\ell,m)$ for the main mode. We covered the parameter space by varying the free parameters in the following way: the projected rotational velocity $v \sin i$ from 28 to 32 km\,s$^{-1}$ with a step 1 km\,s$^{-1}$, the inclination of the star $i_{\rm{rot}}$ from 1$^{\circ}$ to 90$^{\circ}$ with a step 1$^{\circ}$, the line-profile width $\sigma$ from 1 to 20 km\,s$^{-1}$ with a step 1 km\,s$^{-1}$. For each tested $(\ell, m, i)$, we varied the velocity amplitude $A_p$ from 0.8 $A_p^{*}$ to 1.2 $A_p^{*}$ with a step 0.05 $A_p^{*}$ where $A_p^{*}$ is the value of the velocity amplitude which leads to a theoretical first moment amplitude equal to the observed one.

\begin{figure}
\includegraphics[width=8cm]{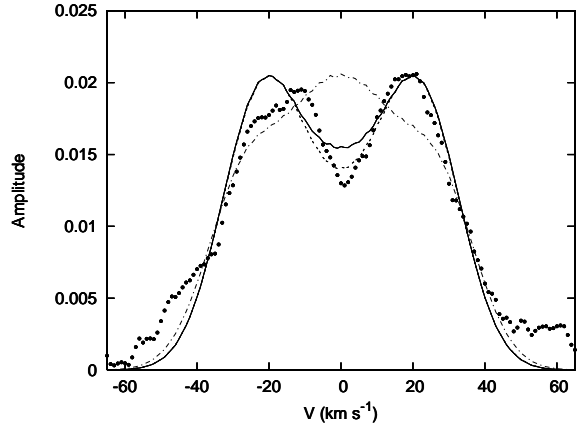}    
\caption{Comparisons between the observed amplitude distribution (dots) for $f_1$ computed for the Si\,III 4553 \AA\ line (dots) and the theoretically computed one for $(\ell_1,m_1) = (2,-1)$ (full line), for $(\ell_1,m_1) = (1,-1)$ (dashed line) and for $(\ell_1,m_1) = (2,-2)$ (dot-dashed line).}  
\label{IPS}  
\end{figure}

In Fig.\,\ref{IPS}, we show the best fit model compared to observation. For a $(2,-1)$ mode, one has $v \sin i$ = 28 km\,s$^{-1}$, $i_{\rm{rot}}$ = 78$^{\circ}$, $\sigma$ =19 km\,s$^{-1}$ and $A_p^1$ = 31.62 km\,s$^{-1}$. For a $(1,-1)$ mode, the best solution is obtained for $v \sin i$ = 29 km\,s$^{-1}$, $i_{\rm{rot}}$ = 70$^{\circ}$, $\sigma$ =19 km\,s$^{-1}$ and $A_p^1$ = 10.20 km\,s$^{-1}$. Finally, for a $(2,-2)$ mode, the parameter values are $v \sin i$ = 32 km\,s$^{-1}$, $i_{\rm{rot}}$ = 90$^{\circ}$, $\sigma$ =19 km\,s$^{-1}$ and $A_p^1$ = 19.29 km\,s$^{-1}$. We point out that the behaviour of these amplitude and phase distributions is the same as the one obtained with the best solutions of the moment method although the values of the continuous parameters are slightly different. Again, we can exclude the $(2,-2)$ solution while we cannot discriminate between a $(2,-1)$ or $(1,-1)$ mode. Since the photometric mode identification in Paper\,I favours $\ell_1 = 2$, we finally conclude that the wavenumbers of the main mode are $(\ell_1,m_1) = (2,-1)$ and consequently the frequencies $f_1 = 7.1160\ \rm{c\,d}^{-1}$, $7.2881\ \rm{c\,d}^{-1}$ and  $f_3 = 7.3697\ \rm{c\,d}^{-1}$ of the rotationally split $\ell = 2$ quintuplet correspond to $m = (-1,1,2)$.

\section[]{Stellar parameters}

By means of the CORALIE \'echelle spectra we derive spectroscopic estimates of the stellar parameters $T_{\rm{eff}}$ and $\log g$ of the primary and tertiary by considering the presence of a B5 companion through the estimated flux ratio derived in Paper\,I. The used spectra were corrected for the orbital velocity obtained in Section\,3 and some 20 spectra with the highest S/N ratio among the 86 CORALIE spectra at our disposal were selected to compute a mean spectrum with an increased S/N ratio and for which the pulsational variability of the primary is cancelled. We used five hydrogen lines ($H_\alpha$, $H_\beta$, $H_\gamma$, $H_\delta$, $H_\epsilon$), six He\,I lines (4026, 4387, 4471, 4713, 4992 and 6678 \AA), two Si\,II doublets (4128-4131 and 5041-5056 \AA), two Si\,III triplets (4553-4568-4575 and 4813-4819-4829 \AA) and two additional Si\,III lines (4716 and 5740 \AA).

These observed spectral lines were compared with theoretical ones for both the primary and tertiary computed for spherically symmetric non-local thermodynamic equilibrium (NLTE) atmosphere models by means of the FASTWIND code in the version of Puls et al.\,\citet{puls}. The main input parameters to obtain the theoretical profile are $T_{\rm{eff}}$, $\log g$, the microturbulent velocity and the particle number ratios $n$(He)/$n$(H), $n$(Si)/$n$(Si$_\odot$). The gravity and temperature are most constrained by respectively the wings of the hydrogen lines and the ionization balance between Si\,II and \,III. 

\begin{figure*}  
\includegraphics[width=12cm,angle=90]{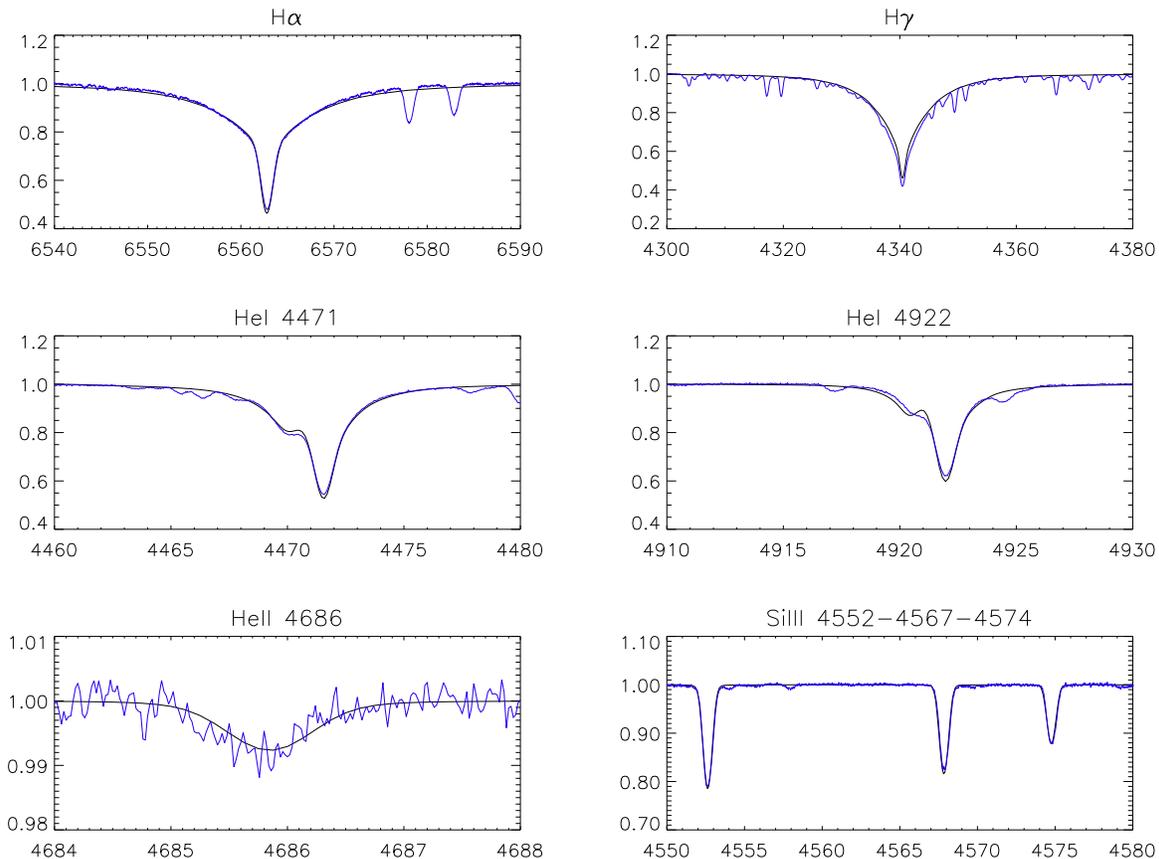}
\caption{Selection of observational hydrogen, helium and silicon line profiles together with their best fits obtained for $T_{\rm{eff}} = 24000 \pm 1000$ K, $\log g = 4.1 \pm 0.1$ for the primary and $T_{\rm{eff}} = 18500 \pm 1000$ K, $\log g = 4.0 \pm 0.1$ for the tertiary and solar He and Si composition.}  
\label{lines}  
\end{figure*}

The best overall fit to the observed line profiles after merging the theoretical profiles for the primary and tertiary occurs for the following parameters. For the primary and the tertiary, one has respectively $T_{\rm{eff}} = 24000 \pm 1000$ K, $\log g = 4.1 \pm 0.1$ and $T_{\rm{eff}} = 18500 \pm 1000$ K, $\log g = 4.0 \pm 0.1$ with solar He and Si abundances. In Fig.\,\ref{lines} we show as examples a selection of observational hydrogen, helium and silicon line profiles together with their best merged line fits. Our results of this spectroscopic determination of stellar parameters is in agreement with the photometric ones derived in Paper\,I.

\section[]{Summary}
Recent thorough studies of a few \mbox{$\beta$ Cephei} stars showed us that probing the internal structure of this kind of stars has become feasible. To this end, very intensive and long-term monitoring is required, as well as accurate determination of many pulsation frequencies together with successful mode identification. For this reason, photometric and spectroscopic multisite campaigns and long-term monitoring of several \mbox{$\beta$ Cephei} pulsators are currently ongoing. \mbox{$\theta$ Ophiuchi} is one of the studied targets for which the observational results are presented in Paper\,I and in this paper. The application of asteroseismic techniques will be described in a following paper.

We can summarize our findings as follows. The 1303 colour photometric three-site data obtained during 77 nights of observation allowed the authors of Paper\,I to detect seven pulsation modes within a narrow frequency interval between 7.116 and 7.973 $\rm{c\,d}^{-1}$. Our less numerous data of 121 spectra were obtained during three years and allowed us to discover a low-mass spectroscopic companion of the B2 primary and to determine an orbital period of some 57 days. The presence of a Speckle B5 companion was already known so that we can conclude that \mbox{$\theta$ Oph} is a triple system. Moreover, we could recover three of the seven pulsation frequencies of the primary observed in photometry by means of a frequency analysis on the radial velocity but also from a two dimensional frequency search on the line profiles themselves.  

The photometric and spectroscopic mode identifications turned out to be complementary and fully consistent with each other. The colour photometry attributed unambiguously the frequency 7.4677 $\rm{c\,d}^{-1}$ to a radial mode, which is corroborated by spectroscopy. Two multiplets are also identified in photometry: one complete $\ell=1$ triplet, which is not observed in the spectroscopic data at our disposal, as well as three components of an $\ell=2$ quintuplet, of which we detected two components in our spectroscopic data. Because this quintuplet is not complete, there are still two possibilities for the $m$-values of the observed $\ell = 2$ modes. This ambiguity was lifted thanks to our spectroscopic mode identification, taking into account the photometric identification of the radial mode. By comparing the observed first three moments of the Si\,III 4553 \AA\ line as well as the observed amplitude and phase variations across this silicon line to the same theoretically computed quantities for several wavenumbers $(\ell,m)$ and for a large grid of continuous parameters, we identified the main mode with frequency 7.1160 $\rm{c\,d}^{-1}$ as a non-axisymmetric $(2,-1)$ mode and consequently we fixed the mode identifications of the whole quintuplet. 

The position of the primary in the Hertzsprung-Russel diagram was also photometrically and spectroscopically determined, leading to compatible stellar parameter values.

These observational results form a good starting point for in-depth seismic modelling of the star. In particular, since its frequency spectrum is so similar to the one of \mbox{V836 Cen}, we have the opportunity to test if the occurence of core overshooting and non-rigid rotation found for that star (Aerts et al.\,\citet{aerts2}, Dupret et al.\,\citet{dupret}) also applies to \mbox{$\theta$ Oph}.

\section*{Acknowledgments}
We thank Dr.\,P.\,Hadrava for making his code FOTEL available to us. We thank Dr.\,G.\,Handler for our fruitful collaboration as well as for comments on a draft version of this paper.

\label{lastpage}

\end{document}